\documentclass[12pt]{article}
\usepackage{graphicx}
\usepackage{amssymb}
\usepackage[english]{babel}
\usepackage{makeidx,verbatim}
\parskip=\medskipamount
\newcommand{\be}{\begin{equation}}
\newcommand{\ei}{\end{equation}}
\begin{document}
\begin{center}\Large The Copenhagen interpretation, and pragmatism\footnote{Contribution to
the Conference on ``Pragmatism \& quantum mechanics'', CREA, \'Ecole Polytechnique \& CNRS, Paris, February 22-23, 2007.}\\
Willem M. de Muynck\\\normalsize
Eindhoven University of Technology,\\
Eindhoven, the Netherlands
\end{center}

\begin{abstract}
In the past both instrumentalism and empiricism have inspired
certain pragmatic elements into the Copenhagen interpretation of
quantum mechanics. The relation of such pragmatisms with the
correspondence principle is discussed. It is argued that neither
Bohr nor Heisenberg did take `correspondence' in one of these
forms, and that it, in particular, was Bohr's classical attitude
which caused him to apply in an inconsistent way his
correspondence principle to the Einstein-Podolsky-Rosen
experiment, thus causing much confusion. It is demonstrated that
an empiricist pragmatism is conducive to an explanation of
violation of the Bell inequalities as a consequence of
`complementarity' in the sense of `mutual disturbance in a joint
nonideal measurement of incompatible observables' rather than as
being caused by `nonlocal influences'.
\end{abstract}

\section{Introduction}\label{sec1}
Let me first apologize for not being a philosopher, and, hence,
not knowing precisely what in philosophic discourse is meant by
pragmatism. For this reason in my discussion of the Copenhagen
interpretation I will start from a physicist's notion of
pragmatism in the sense of `employing theory in a
\textit{practical} way so as to obtain useful results, without
bothering too much about correspondence of theoretical entities
with reality'. An \textit{instrumentalist} interpretation of
quantum mechanics, in which the mathematical formalism is
considered to be merely an instrument for calculating
probabilities of measurement results, seems to come close to this.
As is well known, Bohr had an instrumentalist interpretation of
the quantum mechanical wave function, to which he attributed a
purely symbolic meaning only.

Instrumentalism is liable to criticism. Quantum mechanics is not
pure mathematics. Like for every \textit{physical} theory, some
connection must be established with the physical reality it
purports to describe. If this is omitted, doors are wide open for
all kinds of confusions, as, indeed, have plagued the Copenhagen
interpretation. Thus, an instrumentalist interpretation does not
make a choice between the following two possible meanings of a
`quantum mechanical measurement result', mathematically
represented by an eigenvalue $a_m$ of an Hermitian operator $A$:
i) a property of the microscopic object, ii) a pointer position of
a measuring instrument. Defining an `interpretation of a physical
theory' as a `mapping of the mathematical formalism of that theory
into reality', I take it as one of my tasks to demonstrate that
instrumentalism has caused failure of the Copenhagen
interpretation to be a sound interpretation of quantum mechanics.
As far as instrumentalism is pragmatic, this might be taken as a
criticism of pragmatism.

In the literature also other reasons than instrumentalism can be
found for attributing pragmatism to the Copenhagen interpretation.
Thus, according to Stapp \cite{StapponCIQM} the Copenhagen
interpretation is pragmatic in the sense that quantum mechanical
probabilities (or relative frequencies)
 \begin{equation}\label{1.1} p_m = \langle\psi|E_m|\psi\rangle \mbox{ \rm or }Tr \rho E_m\end{equation}
are not considered as referring to the \textit{microscopic} object
itself, but to \textit{macroscopic} events obtained when a
preparation procedure (represented by wave function $\psi$ or
density operator $\rho$) is followed by a measurement procedure
(represented by the Hermitian operator $A=\sum_m a_m E_m,\;E_m$
projection operators). It, indeed, might be felt to be pragmatic
to forgo all metaphysical discussion by being satisfied with
describing `just the phenomena' rather than `microscopic reality
itself' (e.g. by completely ignoring the relation between a click
in a Geiger counter and the microscopic object causing that click
to occur). In this sense correspondence does not seem to be
inconsistent with pragmatism.

As we will see, at least Bohr's authoritative version of the
Copenhagen interpretation is not in agreement with Stapp's
characterization. Nevertheless, the latter is attractive due to
its exclusive reliance on empirically accessible data, as well as
because of its faithful representation of actual practice in
experimental physics. For this reason it was recently proposed by
the author \cite{dM2004} to amend the Copenhagen interpretation in
such a way as to define a new interpretation, referred to as
\textit{neo}-Copenhagen interpretation, in which the empiricist
kind of pragmatism attributed by Stapp to the Copenhagen
interpretation is actually satisfied.

\begin{figure}[t]
\leavevmode \centerline{
\includegraphics[width=10cm]{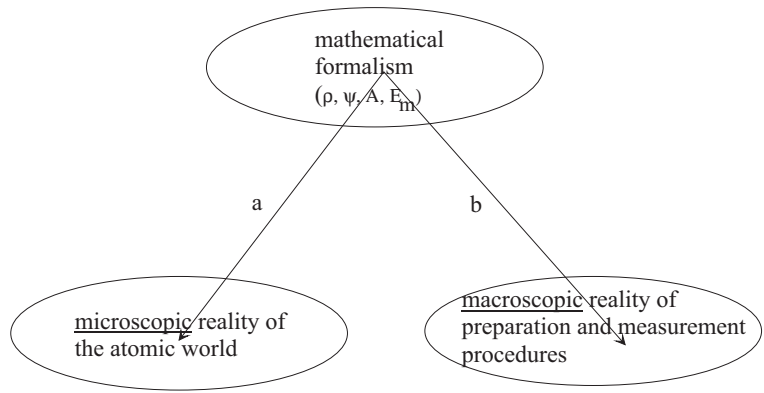}
}\caption{\em\small Realist (a) and empiricist (b) interpretations
of the mathematical formalism of quantum mechanics.} \label{fig1}
\end{figure}

I prefer to refer to the characterization of the Copenhagen
attitude proposed by Stapp as an `\textit{empiricist
interpretation} of the mathematical formalism of quantum
mechanics', to be contrasted with a `\textit{realist}
interpretation' in which the mathematical entities of that
formalism  are thought to be mapped into \textit{microscopic}
reality. Thus,

\noindent \underline{Realist interpretation} (fig.~\ref{fig1}a):\\
Quantum mechanical observable $A$ (in particular, its eigenvalues
$a_m$), wave function $\psi$ and density operator $\rho$ refer to
\textit{properties} of the microscopic object.

\noindent\underline{Empiricist interpretation} (fig.~\ref{fig1}b):\\
Quantum mechanical observable $A$ and its values $a_m$ are
\textit{labels} of a \textit{measurement procedure}, and of the
\textit{pointer positions} of the measuring instrument,
respectively; wave function $\psi$ and density operator $\rho$ are
\textit{labels} of \textit{preparation procedures}.

An interpretation along Stapp's empiricist lines is very well
possible (e.g. de Muynck \cite{dM2002}), and certainly has
adherents among Copenhagen philosophers and physicists. However,
neither Bohr nor Heisenberg --the founding fathers of the
Copenhagen interpretation-- were pragmatic in this empiricist
sense. It is true that the measurement arrangement plays an
important role in their quantum philosophies. But not in the
empiricist sense given above. Notwithstanding their empiricist
terminology in which is referred to a measurement as a `quantum
\textit{phenomenon}', for both Bohr and Heisenberg a quantum
mechanical observable refers to a property of the
\textit{microscopic} object, \textit{not} to a pointer of a
measuring instrument. As far as such instruments play a role in
their reasonings, they serve to \textit{define} (Bohr) or to
\textit{actualize} (Heisenberg) properties of the
\textit{microscopic} object.

In the following sections the three basic issues of the Copenhagen
interpretation, viz. correspondence, completeness, and
complementarity will be discussed from the point of view of an
empiricist interpretation of the quantum mechanical formalism.
This interpretation will also be applied to elucidate the problem
of the Bell inequalities.

\section{Correspondence}\label{sec2}
I will restrict myself to the mature form of Bohr's correspondence
principle, developed after the mathematical formalism of quantum
mechanics had largely been established (referred to in
\cite{dM2002} as the \textit{strong} form of correspondence, so as
to distinguish it from the \textit{weak} form requiring existence
of a classical limit). This principle can be characterized by the
following two requirements: i) experimental arrangement and
measurement results have to be described in classical terms; ii) a
quantum mechanical observable is exclusively defined within the
context of the measurement serving to measure that observable.

Both points are liable to criticism. The first point has its
origin in the logical positivist ideal of basing a theory on
theory-independent observational data, so as to evade the vicious
circularity caused by a dependence of the measurement on the very
theory it is testing. For this reason, according to Bohr quantum
mechanical measurement results should be expressed in terms of an
independently tested theory, viz. classical mechanics. Nowadays we
are convinced, however, that the requirement of
theory-independence of observation statements cannot be met.
Indeed, granting Bohr the necessity that a quantum mechanical
measuring instrument have a \textit{macroscopic} part, viz. a
pointer, of which the position can be described by classical
mechanics, we have become aware of the fact that it must also have
a part which is sensitive to the \textit{microscopic} information
that has to be transmitted from the microscopic object to the
measuring instrument (in order to be finally amplified to
macroscopic dimensions). The microscopic process of information
transfer (often called the pre-measurement) is actually the most
important part of the measurement process; it should be described
by quantum mechanics.

As to the second point the Copenhagen interpretation is liable to
criticism because, notwithstanding the macroscopic part of the
measuring instrument is playing an important role (as is seen from
the first point), it completely ignores the measuring instrument
\textit{as a dynamically involved object}. In their discussions of
the `thought experiments' Bohr and Heisenberg are not so much
interested in the final pointer position of the measuring
instrument; their interest is rather directed toward properties of
the \textit{microscopic} object (discussed in classical terms) as
these properties are \textit{within the context of the
measurement} (Bohr) or even as they are \textit{after} the
measurement has been completed (Heisenberg) (e.g. de Muynck
\cite{dM2002}, chapt.~4). Indeed, the Copenhagen interpretation of
quantum mechanical observables is not an empiricist one as defined
in sect.~\ref{sec1}; on the contrary, the interpretation is a
realist one, be it of a contextualistic kind: a quantum mechanical
measurement result is thought to be a property of the microscopic
object, be it \textit{not} an \textit{objective} property
possessed already \textit{before} the measurement (Einstein), but
only well-defined \textit{within the context of the measurement}
(Bohr) or \textit{after} the measurement (Heisenberg).

\begin{figure}[t]
\leavevmode \centerline{
\includegraphics[height=4.7cm]{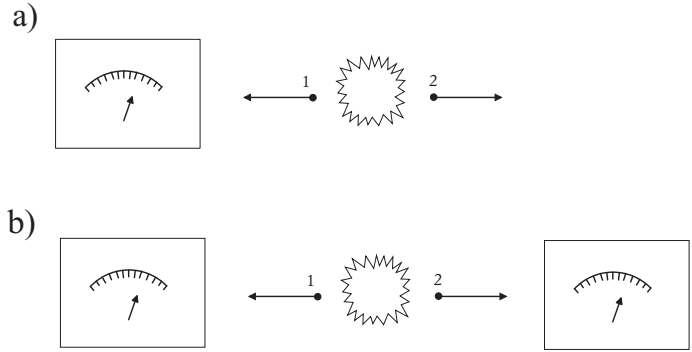}
}\caption{\em\small a) EPR experiment; b) EPR-Bell
experiment.}\label{fig2}
\end{figure}

That the Copenhagen interpretation of Bohr and Heisenberg is not
empiricist but contextualistic-realist can be concluded from many
of their utterances; it can also most clearly be seen from Bohr's
reaction \cite{Bohr35} to the Einstein-Podolsky-Rosen (EPR)
experiment \cite{EPR}, proposed to challenge the Copenhagen
completeness thesis. This experiment was presented by EPR as a
\textit{measurement} of a property of particle $2$ (cf.
fig.~\ref{fig2}a), \textit{without letting this particle interact
with a measuring instrument}.
It is remarkable that it was accepted by Bohr as such a
measurement, because the possibility of measuring a property of
particle $2$ by measuring a property of particle $1$ must be based
on the assumption that the correlation of these properties is
well-defined. But, according to Bohr's correspondence principle
such a correlation would be well-defined only within the context
of a measurement of that correlation. However, a measurement of
the correlation would require a measurement arrangement of the
type depicted in fig.~\ref{fig2}b, referred to as an EPR-Bell
experiment because such measurements have been exploited to test
the Bell inequalities (e.g. the experiments carried out by Aspect
and coworkers \cite{Aspect})\footnote{It should be noted that
these experiments are often referred to as EPR experiments too,
thus perpetuating the confusion originating with Einstein and
Bohr.}. So, the fact that Bohr accepted the EPR experiment as a
measurement of a property of particle $2$ actually amounted to an
inconsistent application by Bohr of his correspondence principle.
Evidently, Bohr did not recognize `correlation' as a quantum
mechanical observable, to be well-defined only within the context
of a correlation measurement. Most probably this oversight is a
consequence of his \textit{realist} interpretation of such
physical quantities. Indeed, the Einstein-Bohr controversy, having
its apotheosis in their discussion on the EPR paper \cite{EPR}, is
about the question of whether such a \textit{realist}
interpretation of quantum mechanical observables can be
\textit{objectivistic} (Einstein: observables are thought to be
independent of the `observer \textit{including his measuring
instruments}'), or must be \textit{contextualistic} (Bohr,
Heisenberg: observables are thought to be dependent on the
experimental arrangement).

In quantum mechanics textbooks `measurement' is treated
axiomatically; the measuring instrument is not dealt with in an
explicit way. The quantum mechanical theory is treated as a
description of \textit{microscopic} reality. The wave function is
thought not to refer to a preparation procedure, but to the
\textit{result} of such a procedure. By the same token a
measurement result is not considered as referring to a property of
a measuring instrument, but to a property of the microscopic
object. Hence textbooks entertain a \textit{realist}
interpretation of the mathematical formalism of quantum mechanics,
as do most quantum mechanical publications in scientific journals.
Nowadays it is increasingly realized that such a realist
interpretation cannot be taken in Einstein's (or textbook)
objectivistic sense, but should at least be contextualistic to be
able to cope with no-go theorems like the Kochen-Specker theorem
and the Bell inequalities.

Bohr's awareness of the important role of measurement in the
interpretation of quantum mechanics certainly earns him the
victory, widely attributed to him, over Einstein's objectivistic
realism. However, we will see that a contextualistic realism alone
is not able to solve all problems of interpretation: it will be
necessary to accept the still weaker empiricist interpretation. As
we shall see in sections~\ref{sec4} and \ref{sec5}, is the
empiricist interpretation able to deal in a satisfactory way with
Bohr's notion of complementarity (which is a key notion of quantum
mechanics), as well as with the Bell inequalities. By relying on
the empiricist interpretation it is not only possible to avoid the
ambiguities stemming from an \textit{instrumentalist}
interpretation in which no choice is made between a (realist)
property of the microscopic object and an (empiricist) pointer
position of the measuring instrument (as is evidenced by the
widely practiced confounding of EPR and EPR-Bell experiments), but
it is also possible to evade dubious consequences of Bohr's
contextualistic-realist interpretation.

Returning to the question of pragmatism, we can learn from this
discussion of the Copenhagen correspondence principle that i) the
Copenhagen interpretation is neither pragmatic in the
instrumentalist sense (due to its reliance on `correspondence'),
ii) nor is it pragmatic in the empiricist sense (due to its
`\textit{realist} correspondence'). There undoubtedly is a certain
pragmatic tendency in Bohr's contextualism, to the effect that one
should be satisfied with `knowledge about the microscopic object
\textit{as it is within the context of a measurement}' (since, due
to the disturbing influence of measurement, \textit{objective}
knowledge in Einstein's sense is thought to be an unattainable
ideal). However, this pragmatism should not be confused with the
empiricist pragmatism defined above. Although Bohr's notion of
`quantum phenomenon' sounds deceivingly empiricist, it should in
general not be equated with a `measurement phenomenon' like a
flash on a screen or a click in a counter. By sticking too much to
the realist interpretation of classical mechanics (be it amended
in a contextualistic sense) Bohr has not been able to benefit
sufficiently from the possibilities the empiricist pragmatism has
to offer. In particular, the empiricism of the neo-Copenhagen
interpretation solves the problem that quantum mechanical
measurement results did not exist before the measurement, but must
come into being during that measurement: this is trivially
satisfied if measurement results correspond to \textit{final}
pointer positions of measuring instruments.

\section{Completeness}\label{sec3}

We have to distinguish two senses of completeness, viz.
`completeness in a wider sense' and `completeness in a restricted
 sense', the first one turning around the question
`Are hidden variables possible?' whereas the second regards the
question `Does quantum mechanics describe all measurements
possible within the domain of atomic physics?'

\subsection{Completeness in a wider sense}\label{sec3.1}
It is well-known that the EPR paper \cite{EPR} was meant to
challenge the Copenhagen completeness claim taken in a wider
sense, by attempting to prove that quantum mechanical observables
can play the roles of hidden variables, an objectively possessed
value $a_m$ being assumed to be simultaneously attributable to
each observable in the initial state of the microscopic object
(so-called `element of physical reality'). If possible this would
support Einstein's \textit{ensemble} interpretation of the quantum
mechanical wave function as against the Copenhagen
\textit{individual-particle} interpretation.

On the basis of the Kochen-Specker theorem \cite{KS}, proven many
years after the EPR paper was published, we can conclude that
Einstein's idea (that quantum mechanical measurement results $a_m$
can be looked upon as objective properties of the microscopic
object, possessed independently of the measurement, as is the case
in classical mechanics) is not consistent with the mathematical
formalism of quantum mechanics. At the time of the EPR discussion,
however, Einstein could maintain his claim, and even strengthen it
\cite{Einst48}, by remarking that Bohr's reproach \cite{Bohr35} of
ambiguity of the notion of `element of physical reality' actually
implied an unphysical consequence of `nonlocality' which could be
traded off against his conclusion of `incompleteness': according
to Einstein \textit{locality} could be maintained if
`\textit{in}completeness' of quantum mechanics is accepted; only
if quantum mechanics were assumed to be complete, would
nonlocality become an issue.

The festival of confusions involved in the EPR problem as
discussed by Bohr and Einstein cannot be dealt with here in its
entirety (see e.g. de Muynck \cite{dM2002}, sect.~6.5). Suffice it
to make two remarks. First, the possibility of an `ensemble
interpretation' is not at all thwarted by the failure of
Einstein's `elements of physical reality' to be represented by
quantum mechanical measurement results (e.g. Guy and Deltete
\cite{GuyDel90}). There is no reason to exclude ensembles in which
the EPR `elements of physical reality' are represented by
\textit{sub}quantum properties (as is done in hidden-variables
theories considered in later derivations of the Bell
inequalities), rather than by quantum mechanical measurement
results; particularly so if these latter refer to pointer
positions of measuring instruments, as is the case in an
\textit{empiricist} interpretation. It seems that the possibility
of subquantum theories, made respectable by John Bell's opening
towards experimental testing, has considerably changed physicist's
attitudes. Whereas a Copenhagen physicist A.D. 1935 would have
answered the question of `completeness of quantum mechanics in the
wider sense' in the following vein: ``Quantum mechanics is
complete; there are no hidden variables'', would a neo-Copenhagen
physicist A.D. 2007 give the converse answer that ``Quantum
mechanics is incomplete; hidden variables theories may be
necessary to describe reality behind the phenomena.'' The latter
would base his judgment on i) Bell's disproof \cite{Bell66} of the
adequacy of von Neumann's `no go' theorem (\cite{vN32},
sect.~IV.2), ii) experimental evidence (e.g. \cite{Moell})
provided by interference (`which way') experiments to the effect
that an interference pattern, described by the wave function, is
gradually built up out of local impacts of an \textit{ensemble} of
individual particles.

My second remark regards the nonlocality issue. For Bell the
existence of Bohm's causal interpretation of quantum mechanics,
considered to be a hidden-variables theory, and, hence, believed
to be a palpable disproof of von Neumann's and other's `no go'
theorems, was a third reason for accepting the possibility of
subquantum theories. Moreover, the nonlocality of Bohm's theory
was reason for him to believe that the underlying reality
described by such theories should exhibit nonlocal features, thus
corroborating the nonlocality allegedly already present in the EPR
experiment. Since it is questionable whether Bohm's interpretation
of quantum mechanics does allow an interpretation as a
hidden-variables theory (e.g. de Muynck \cite{dM2002},
sect.~10.3), this third issue should, however, presumably be seen
as a late act in the EPR festival of confusions.

It should finally be mentioned here that neither Bohr nor
Heisenberg believed quantum mechanics to be `complete in a wider
sense'. We will return to the nonlocality issue in
sect.~\ref{sec5}, where it will become evident that, like the
ensemble issue, also the nonlocality issue is a consequence of
sticking too much to a \textit{realist} interpretation of the
notion of a quantum mechanical observable.

\subsection{Completeness in a restricted sense}\label{sec3.2}
\begin{figure}[t] \leavevmode \centerline{
\includegraphics[height=3.5cm]{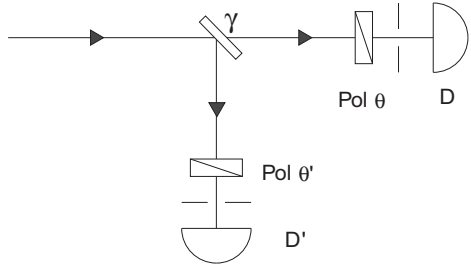}
} \caption{\em \small Which-way polarization measurement of a
photon.} \label{fig4}
\end{figure}

The answer to the question of `whether quantum mechanics describes
all measurements possible within the domain of atomic physics' is
dependent on whether one restricts oneself to the
\textit{standard} formalism to be found in quantum mechanics
textbooks, to the effect that quantum mechanical probabilities
$p_m$ satisfy the Born rule (\ref{1.1}), in which the operators
$E_m$ are projection operators satisfying $E_m^2=E_m$. If this
\textit{standard} quantum mechanics is meant, the answer is
unambiguously ``no''. Indeed, experiments satisfying (\ref{1.1})
do not exhaust all possible quantum mechanical measurements. On
the contrary, most realistic measurements turn out to yield the
more general probabilities
\begin{equation}\label{3.2}
p_m = \langle\psi|M_m|\psi\rangle \mbox{ \rm or } Tr \rho M_m,\;
\sum_m M_m=I, \;M_m\geq O,
\end{equation}
in which the operators $M_m$ need not be projection operators. The
set of operators $\{M_m\}$ defines a `\textit{non}-orthogonal
resolution of the identity', or a `positive operator-valued
measure (POVM)'. Measurements satisfying (\ref{1.1}) constitute a
subset (corresponding to an `orthogonal resolution of the
identity', or `projection-valued measure (PVM)') of the set of
measurements satisfying (\ref{3.2}). For instance, consider the
`which-way polarization measurement of a photon' depicted in
fig.~\ref{fig4}.
When a photon impinges on a semitransparent mirror it has
probability $\gamma$ to be transmitted and probability $1-\gamma$
to be reflected. Hence, the detection probabilities of
polarization detectors D and D$'$ are given by $P_{\rm D} =\gamma
\langle\psi| E_+^\theta|\psi\rangle$ and $P_{\rm D'} =(1-\gamma)
\langle\psi| E_+^{\theta'}|\psi\rangle$, respectively, in which
$E_+^\theta$ and $E_+^{\theta'}$ are projection operators of the
corresponding standard polarization observables. It is evident
that the detection probabilities are not represented by these
latter projection operators, but by the positive operators $\gamma
E_+^\theta$ and $(1-\gamma) E_+^{\theta'}$, respectively. Together
with the operator $I-\gamma E_+^\theta - (1-\gamma) E_+^{\theta'}$
(representing the probability that a photon is absorbed in one of
the analyzers), these operators define a POVM as employed in
(\ref{3.2}). Only in the exceptional cases in which $\gamma$ is
either $1$ or $0$ this POVM reduces to a PVM. Evidently, out of
all possible experiments of this kind standard quantum mechanics
merely covers a set of measure zero.

Taking into account only the experiments corresponding to the
limiting values $\gamma= 1$ and $\gamma=0$ (as is done in the
Copenhagen interpretation as well as in most other interpretations
of quantum mechanics) implies that \textit{most} empirical
information, to be obtained by means of measurements of the type
depicted in fig.~\ref{fig4}, is ignored. It seems to me that from
a pragmatic point of view this must be utterly undesirable.
Indeed, as will be illustrated in the following sections, reliance
on too restricted a theory (viz. the \textit{standard} formalism
of quantum mechanics, describing only measurements corresponding
to PVMs) has had a very large and probably rather dubious effect
on the way physicists have been thinking about how reality is
constituted. Although within the context of discovery it may have
been advantageous to stick to the standard formalism of quantum
mechanics in order to be able to obtain fast results by applying a
relatively simple mathematical formalism, it seems that within the
context of justification such pragmatism may have a restraining
influence by causing stagnation due to misconceptions based on too
scant empirical evidence.

\section{Complementarity}\label{sec4}
Complementarity, as discussed within the Copenhagen
interpretation, is about the (im)possibility of simultaneously
measuring incompatible (standard) observables corresponding to
noncommuting Hermitian operators. Allegedly, such measurements are
impossible due to the disturbing influence of a measurement of
observable $A$, say, on the measurement results to be obtained of
the other observable $B$ if $[A,B]_- \neq O$. By considering
`thought experiments' like double slit experiments and the
$\gamma$-microscope, such a disturbance was unambiguously
demonstrated to occur. This seemed to be corroborated by the
possibility to derive from the mathematical formalism of quantum
mechanics the well-known \textit{Heisenberg uncertainty relation}
 \begin{equation}\label{4.1}\Delta A \Delta B
\geq \frac{1}{2} | \langle
\psi|[A,B]_-|\psi\rangle|,\end{equation} in which $\Delta A$ and
$\Delta B$ are standard deviations in state $\psi$. For a long
time the Heisenberg uncertainty relation (\ref{4.1}) was
considered to be an expression of the mutual disturbance of
measurement results of \textit{incompatible} observables $A$ and
$B$ if these are measured simultaneously.

It lasted about $40$ years before Ballentine \cite{Bal70} opposed
this view by noting that the Heisenberg uncertainty relation
(\ref{4.1}) can be tested by \textit{separate} measurements of
observables $A$ and $B$. Hence, this relation does not at all
refer to a \textit{simultaneous} measurement of $A$ and $B$.
Evidently, by equating the inequalities they had derived for a
number of `thought experiments' with the theoretical inequality
(\ref{4.1}) at hand, Bohr and Heisenberg had jumped to conclusions
as regards the physical meaning of (\ref{4.1}).  Indeed, a
reasonable interpretation of (\ref{4.1}) could be that of a
property of the \textit{preparation procedure} of the
\textit{ensemble} represented by $\psi$ (rather than a property of
a measurement), more or less in Einstein's sense, to the effect
that it is impossible to \textit{prepare} an ensemble for which
the physical quantities $A$ and $B$ are both dispersionless.

This does not imply, however, that the Bohr-Heisenberg idea of
mutual disturbance in a simultaneous measurement of incompatible
observables would be incorrect. Nowadays we are able to carry out
realistic experiments, to be interpreted as joint (nonideal)
measurements of incompatible standard observables, experimentally
exhibiting such a disturbance (e.g. de Muynck \cite{dM2002},
chapt.~8). But a quantum mechanical description of these
experiments requires the generalization of the mathematical
formalism of quantum mechanics referred to in sect.~\ref{sec3.2},
encompassing measurements labelled by POVMs rather than by PVMs.
In this section the measurement depicted in fig.~\ref{fig4} will
be discussed as an example of such a joint nonideal measurement.
By registering for each individual incoming photon the reactions
of \textit{both} detectors D and D$'$, the experiment gives
occasion to define a \textit{joint} detection probability
$p_{mn},\;m,n =+$ or $-$, in which $p_{++} =0$, $p_{+-}=p_{\rm
D}'$, $p_{-+}=p_{\rm D'}$, and $p_{--}$ is the probability of a
photon being absorbed in one of the analyzers. Writing
\begin{equation}p_{mn}= \langle\psi| M_{mn}^\gamma|\psi\rangle\label{4.0}\end{equation}
 it follows
that the POVM can be represented in the following bivariate form:
\begin{equation} (M^\gamma_{mn})= \left(
\begin{array}{cc}
O & \gamma  E^\theta_+\\
(1-\gamma) E^{\theta'}_+ &  1-\gamma E^\theta_+ - (1 - \gamma)
 E^{\theta'}_+
\end{array} \right).\label{4.2}\end{equation}
By taking marginals it is possible to find the detection
probabilities of each of detectors $\rm D$ and $\rm D'$
separately, allowing to interpret the measurement as a joint
nonideal measurement of the incompatible standard polarization
observables in directions $\theta$ and $\theta'$, represented by
the PVMs $\{E^\theta_+,E^\theta_-\}$ and
$\{E^{\theta'}_+,E^{\theta'}_-\}$, respectively. We find, for
incoming wave function $\psi$:
\begin{equation} \begin{array}{l}
\mbox{\rm detector D}:\; \left(\begin{array}{c}\sum_n
p_{+n}\\\sum_n p_{-n}\end{array}\right) = \left(\begin{array}{cc}
\gamma & 0\\1-\gamma& 1\end{array}\right)
\left(\begin{array}{c} \langle\psi| E^\theta_+|\psi\rangle\\\langle\psi| E^\theta_-|\psi\rangle\end{array}\right),\\
\\
\mbox{\rm detector D}':\;\left(\begin{array}{c}\sum_m
p_{m+}\\\sum_m p_{m-}\end{array}\right) = \left(\begin{array}{cc}
1-\gamma & 0\\\gamma& 1\end{array}\right) \left(\begin{array}{c}
\langle\psi| E^{\theta'}_+|\psi\rangle\\\langle\psi|
E^{\theta'}_-|\psi\rangle\end{array}\right).
\end{array}\label{4.3}\end{equation}
In these expressions the nonideality matrices \tiny
$\left(\begin{array}{cc} \gamma & 0\\1-\gamma&
1\end{array}\right)$ \normalsize and \tiny
$\left(\begin{array}{cc} 1-\gamma & 0\\\gamma&
1\end{array}\right)$ \normalsize represent the nonidealities in
the determination of the probabilities of the standard
polarization observables in directions $\theta$ and $\theta'$,
respectively, yielded by the present experiment.

\begin{figure}[t] \leavevmode \centerline{
\includegraphics[height=6cm]{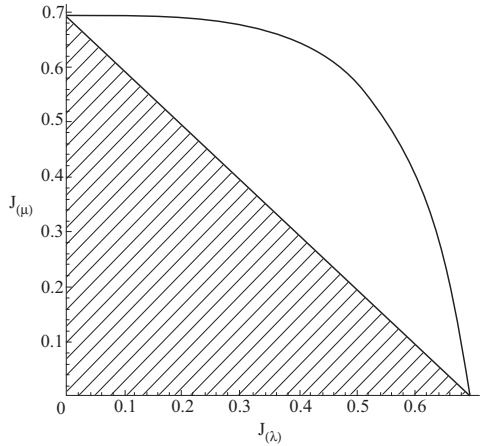}
} \caption{\em \small Parametric plot of $J_{(\protect\lambda) }$
versus $J_{(\protect\mu) }$ as a function of $\gamma$.}
\label{fig5}
\end{figure}

It is important to notice the complementary behaviour of the two
nonideality matrices given above: the quality of the information
yielded by one marginal probability distribution increases as that
of the other marginal probability distribution decreases by
changing the value of $\gamma$. Indeed, for $\gamma =1$
information on the standard polarization observable in direction
$\theta$ is ideal whereas that on the standard polarization
observable in direction $\theta'$ is maximally nonideal. For
$\gamma=0$ the opposite holds. For $0<\gamma<1$ information on
both standard observables is nonideal to a certain extent.
Denoting the two nonideality matrices that are involved by
$(\lambda_{mm'})$ and $(\mu_{nn'})$, respectively, and taking the
average row entropy
\[J_{(\lambda)} = - \frac{1}{N} \sum_{mm'} \lambda_{mm'} \ln \frac{\lambda_{mm'}}
{\sum_{m'} \lambda_{mm'}}\]
 as a measure of the nonideality expressed by the matrix $(\lambda_{mm'})$
(and analogously for $(\mu_{nn'})$), by Martens \cite{MadM90} an
inequality was derived, for the present measurement reading
\begin{equation}\label{3.1} J_{(\lambda)} + J_{(\mu)} \geq -\ln \{\max_{mn} Tr
E^{\theta}_m E^{\theta'}_n\}.\end{equation}
 In fig.~\ref{fig5}
the curved line is a parametric plot of $J_{(\protect\lambda) }$
versus $J_{(\protect\mu) }$ as a function of $\gamma$. The shaded
area contains the values of $J_{(\protect\lambda)} $ and
$J_{(\protect\mu) }$ forbidden by the Martens inequality
(\ref{3.1}).

Contrary to the Heisenberg inequality (\ref{4.1}) the Martens
inequality (\ref{3.1}) yields a faithful representation of the
idea of mutual disturbance in a simultaneous measurement of
incompatible standard observables, as found in the early
discussions on the `thought experiments'. In particular, it can be
seen that the Martens inequality, being completely independent of
the initial state vector or density operator, refers to the
\textit{measurement procedure} alone, and can be understood as a
consequence of the mutual exclusiveness of measurement
arrangements for measuring incompatible standard observables.
Hence, the Martens inequality does execute the function
erroneously attributed to the Heisenberg inequality. Evidently,
the physical intuitions of Bohr and Heisenberg were sufficiently
adequate. However, they lacked the mathematical formalism (i.c.
the formalism of positive operator-valued measures) necessary for
a mathematical expression of these intuitions. Due to the
restriction of the formalism to standard observables, and the
corresponding restriction of the domain of application of quantum
mechanics, they were not able to see that the notion of
complementarity actually consists of two different forms, viz. one
for \textit{preparation} and one for \textit{measurement}, the
former being represented by the Heisenberg uncertainty relation,
the latter by the Martens inequality.

This clearly illustrates the risk, already pointed at in
sect.~\ref{sec3.2}, that a too narrow scope of the domain of
interest may lead to confusion. Moreover, the example of the POVM
(\ref{4.2}) demonstrates that a restriction of the empiricist
interpretation to the \textit{standard} formalism may generate the
erroneous idea that an ideal measurement of a standard observable
does not yield any information on an incompatible one. As a matter
of fact, it is easy to verify that from (\ref{4.3}) we find for
$\gamma=1$ (i.e. for the ideal measurement of the standard
polarization observable in direction $\theta$) that $\sum_m
p_{m-}=1$. This implies that the measurement result `$-$' can be
attributed \textit{with certainty} to the standard polarization
observable in direction $\theta'$. Evidently the maxim
`Unperformed experiments have no results' (e.g. Peres
\cite{Peres78}), allegedly making it possible to deny the
existence of the value of the polarization observable in direction
$\theta'$ if the polarization observable in direction $\theta$ is
actually measured, is too shallow. The simultaneous existence of
measurement results for the standard polarization observables in
directions $\theta$ and $\theta'$ will be used in the next section
to analyze the problem of the Bell inequalities.

\section{Bell inequalities}\label{sec5}
As mentioned in sect.~\ref{sec3.1} the EPR experiment has induced
the idea that the reality described by quantum mechanics should
have some feature of nonlocality. On the basis of this assumption
inequalities were derived by Bell \cite{Bell64}, allegedly to be
violated only in case of nonlocality. Bell's expectation is still
widely thought to be corroborated by the EPR-Bell experiments
performed by Aspect and coworkers \cite{Aspect}, in which a
violation of the Bell inequalities was experimentally found. In
this section it is demonstrated that this conclusion, too, is
based on the restricted view of quantum mechanics discussed in
sect.~\ref{sec3.2}, and, hence, too shallow.

Let us first see why it is rather improbable that there is any
causal relation between violation of the Bell inequalities and
nonlocality. As a matter of fact, violation of the Bell
inequalities is \textit{only} possible if \textit{not} all of the
four standard observables $A_1,B_1,A_2$ and $B_2$ that are
involved, are mutually compatible (since the existence of a
quadrivariate probability distribution, as a consequence of
compatibility, would imply satisfaction of the Bell inequalities).
Hence, violation of the Bell inequalities is a consequence of
\textit{in}compatibility of some of the observables that are
involved. But, as a consequence of the principle of local
commutativity (prescribing commutativity of observables measured
in causally disjoint regions of space-time) is
\textit{in}compatibility a \textit{local} affair. Only
measurements performed in \textit{one and the same region} of
space-time can be incompatible, and, hence, disturb each other so
as to cause violation of the Bell inequalities. This well-known
result is often ignored on the basis that the Bell theorem is not
a theorem of quantum mechanics but a hidden-variables theorem.
However, it would be rather far-fetched to believe that two
different theories, valid in the same experimental domain, would
yield diametrically opposed explanations of the same phenomena. As
will be seen in the following, the explanation of violation of the
Bell inequalities on the basis of \textit{local} disturbances will
be corroborated by the generalization of the mathematical
formalism introduced in sect.~\ref{sec3.2}.

\begin{figure}[t]
\leavevmode \centerline{
\includegraphics[height=4.5cm]{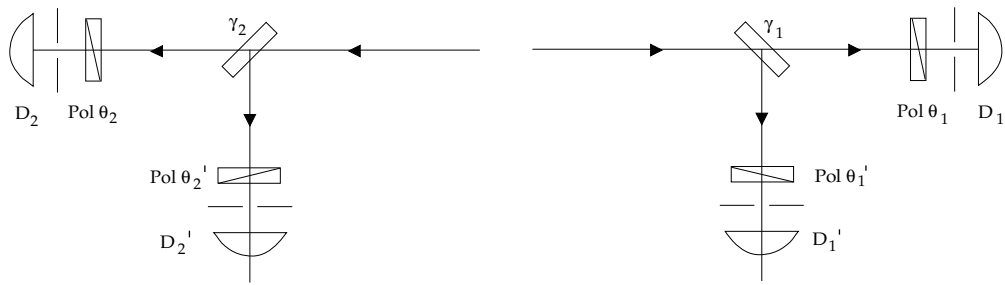}
} \caption{\em \small Generalized EPR-Bell experiment.}
\label{fig6}
\end{figure}

Let us consider the \textit{generalized} Aspect experiment
depicted in fig.~\ref{fig6}.
The Aspect experiments \cite{Aspect} are special cases of this
experiment, in which $(\gamma_1, \gamma_2) =(1,1),(1,0),(0,1)$ or
$(0,0)$. Comparing the experiment with the one depicted in
fig.~\ref{fig4}, and taking into account our discussion in
sect.~\ref{sec4}, we see that the generalized Aspect experiment
can be interpreted as a joint nonideal measurement of four
standard polarization observables. The quadrivariate probability
distribution of the four detectors $\rm D_1$, $\rm D'_1$, $\rm
D_2$ and $\rm D'_2$ is found completely analogously to (\ref{4.0})
as the expectation value of a quadrivariate POVM:
\begin{equation}\label{5.1}
p^{\gamma_1\gamma_2}_{m_{1} n_{1} m_{2} n_{2}} =
\langle\psi|M^{\gamma_1\gamma_2}_{m_{1} n_{1} m_{2}
n_{2}}|\psi\rangle.
\end{equation}
It is also not difficult to see that this POVM is just the direct
product of the bivariate POVMs (\ref{4.2}) of the joint nonideal
measurements performed in each of the arms of the interferometer:
\begin{equation}\label{5.2} M^{\gamma_1\gamma_2}_{m_{1} n_{1} m_{2} n_{2}} =
M^{\gamma_1}_{m_{1} n_{1}} M^{\gamma_2}_{m_{2}
n_{2}}.\end{equation}

It is important to note that, due to the existence of the
quadrivariate probability distribution (\ref{5.1}), its four
bivariate marginals describing correlations between photons $1$
and $2$ do satisfy the Bell inequalities. This, actually, is a
simple consequence of the fact that each measurement on an
individual photon pair yields a \textit{quadruple} of measurement
results, one for each of the four detectors. Note that this holds
true also for each of the special values of the parameters
$\gamma_1$ and $\gamma_2$ employed in the Aspect experiments
\cite{Aspect}. Indeed, violation of the Bell inequalities in these
latter experiments is a consequence of the fact that for the
measurement results of these experiments (constituting octuples
rather than quadruples) no quadruples can be found.

It is often thought that the explanation of the nonexistence of
such quadruples and, hence, of the ensuing violation of the Bell
inequalities in the Aspect experiments, is provided by a nonlocal
influence of the measurement arrangement for photon $1$ on the
measurement results for photon $2$ (and vice versa). As already
put forward, this explanation is not very plausible. There is a
much more plausible explanation, though, viz. the mutual
disturbances of measurement results in each of the arms of the
interferometer \textit{separately}, as a consequence of the fact
that, even for $\gamma_i = 1$ or $0$, the measurements can be
interpreted as joint nonideal measurements of incompatible
observables, which are mutually disturbing. Thus, it is evident
that the measurement result of the standard polarization
observable of photon $1$ in direction $\theta_1$ in case
$\gamma_1=1$ will in general be different from the value found if
$\gamma_1=0$, even if the individual preparations are identical.
Hence, the nonexistence of one single quadruple of measurement
results for the four measurements realized in an Aspect experiment
need not be a consequence of nonlocal influences, but can be
attributed to changing measurement results of detectors D$_1$ and
D$'_1$ if $\gamma_1$ is switched from $1$ to $0$ (and analogously
for D$_2$ and D$'_2$ if $\gamma_2$ is switched). Hence,
disturbances in one arm of the interferometer are caused by
changing the measurement arrangement in that same arm. Once again
it is seen that a natural explanation of a phenomenon may be
overlooked by sticking to a too restrictive experimental and
theoretical domain: in the standard formalism it is not at all
obvious that violation of the Bell inequalities can be seen as a
consequence of complementarity rather than nonlocality.

\section{Conclusions}\label{sec6}
In a physicist's approach to pragmatism, instrumentalist and
empiricist interpretations of the mathematical formalism of
quantum mechanics have been set against the widely used realist
interpretation. It is concluded that, in developing the Copenhagen
interpretation, both Bohr and Heisenberg entertained such a
realist interpretation, be it of a contextualistic kind (as
opposed to Einstein's objectivistic realism). It is argued in
sec.~\ref{sec2} that by this realism Bohr was seduced into
\textit{inconsistently} applying his correspondence principle to
the EPR problem, thus causing much confusion. It is demonstrated
that drawing a clear distinction between the `EPR experiment' and
`EPR-Bell experiments devised for testing the Bell inequalities'
is crucial to lifting the above-mentioned confusions, in
particular the nonlocality conundrum. Thus, in the empiricist
version of pragmatism a consistent application of Bohr's
correspondence principle would not have given rise to any idea of
nonlocal influences.

In sec.~\ref{sec3} two senses of completeness of quantum mechanics
are distinguished. First, `completeness in a wider sense', denying
the possibility of hidden-variables theories, attributed to the
Copenhagen interpretation as one of its main features. In
sec.~\ref{sec3.1} it is argued that, contrary to widespread
belief, it is very well possible that in a pragmatic approach the
Copenhagen `completeness thesis in a wider sense' be replaced by
an \textit{ensemble} interpretation of the wave function. This,
actually, is the default way nowadays quantum mechanical
experiments are dealt with in experimental practice. Here
pragmatism seems to side with scientific progress by cutting off
deliberations regarding currently impracticable tests of
subquantum theories.

With respect to `completeness in a restricted sense' the situation
is different. The question is here whether the domain of
application of quantum mechanics is restricted to measurements
described by the \textit{standard} formalism. As discussed in
sections~\ref{sec3.2}, \ref{sec4} and \ref{sec5}, the answer to
this question is negative since experiments can easily be
performed needing a \textit{generalization} of the mathematical
formalism for their description. It, however, seems that a certain
pragmatism induces physicists to stick to the better-known
standard formalism, even though considerably more insight can be
obtained from the more general experiments described by the
generalized formalism. This is illustrated in sections~\ref{sec4}
and \ref{sec5} by a discussion of the Copenhagen notion of
`complementarity' and by applying a \textit{generalized} EPR-Bell
experiment to the problem of the Bell inequalities. It is seen
that only the \textit{generalized} formalism mathematically
encompasses the notion of `mutual disturbance in a joint nonideal
measurement of incompatible observables', and that violation of
the Bell inequalities is a consequence of this same \textit{local}
incompatibility rather than being caused by nonlocal influences.


\end{document}